\documentclass[12pt]{article}
\usepackage{amssymb}
\def\beq{\begin{equation}}
\def\eeq{\end{equation}}
\usepackage[all,2cell,dvips]{xy}
\def\IR{\relax{\rm I\kern -.18em R}}

\begin{document}
\title{ New non-local SUSY KdV conservation laws from a recursive gradient algorithm }
\author{ \Large S. Andrea*, A. Restuccia**, A. Sotomayor***}
\maketitle{\centerline {*Departamento de Matem\'{a}ticas,}}
\maketitle{\centerline{**Departamento de F\'{\i}sica}}
\maketitle{\centerline{Universidad Sim\'on Bol\'{\i}var}}
\maketitle{\centerline{***Departamento de Ciencias B\'{a}sicas}}
\maketitle{\centerline{Unexpo, Luis Caballero Mej\'{\i}as }}
\maketitle{\centerline{e-mail: sandrea@usb.ve, arestu@usb.ve,
sotomayo81@yahoo.es }}
\begin{abstract} A complete proof of the recursive gradient
approach is presented. It gives a construction of all the
hierarchy structures of $N=1$ Super KdV, including the non-local
one. A precise definition of the ring of superfields involved in
the non-local construction is given. In particular, new non-local
conserved quantities of $N=1$ Super KdV are found.

\end{abstract}

\section{Introduction} KdV equations describe commuting flows in
the space of Schr\"{o}dinger equations,
\[\frac{\partial Q}{\partial t_n}=\left[M_n,Q \right], \] where
$Q=\frac{d^2}{dx^2}+U(x,t_n)$ is a Schr\"{o}dinger operator.

The KdV hierarchy is almost determined by requiring that $[M_n,Q
]$ be a zero order differential operator. Besides their relevance
as an integrable system, KdV equations are directly related to
two-dimensional topological gravity and string theory. It was
conjectured by E. Witten \cite{Witten} that the KdV hierarchy
governs the stable intersection theory on the moduli spaces of
Riemann surfaces. A generalization of that conjecture considers a
Riemann surface $\Sigma$ together with a holomorphic map of
$\Sigma$ to a fixed complex manifold \cite{Witten}. This
holomorphic immersion naturally occurs in the formulation of
$D=11$ Supermembranes with central charges
\cite{Restuccia1,Restuccia2,Restuccia3} which in turn may be
formulated as a noncommutative gauge theory \cite{Restuccia4}.

A supersymmetric extension of the KdV equations was introduced in
\cite{Manin} and independently in \cite{MM,M}, where a detailed
analysis of the system was performed. For a review see
\cite{Mathieus}.

In the same way that the KdV equation is related to the
Schr\"{o}dinger operator of quantum mechanics, the supersymmetric
KdV (SKdV) equations are related to supersymmetric quantum
mechanics. In \cite{Andrea2} it was shown that the Green's
function of the SUSY quantum operator is well defined and that its
asymptotic expansion when $t\rightarrow 0^+$ provides all the SKdV
hierarchy. In \cite{M} a super Gardner transformation was
introduced allowing one to obtain all the known local conserved
quantities of the SKdV equations from a single conserved quantity
of the Super Gardner equation. This super-transformation
generalizes the well-known Gardner transformation for the KdV
equation \cite{Gardner et al}. See also
\cite{Kiselev1,Kiselev2,Andrea1}. An important distinction between
the SKdV and KdV hierarchies is that the former presents non-local
conserved quantities. The earliest non-local conserved quantities
to appear were first presented in \cite{Kersten1} and later in
\cite{Dargis}, where they were obtained from a Lax formulation of
the SKdV hierarchy and generated from the super residue of a
fractional power of the Lax operator. These non-local conserved
quantities are ``fermionic" in distinction to the known ``bosonic"
local ones.

The infinite set of non-local conserved quantities was also
obtained from a single fermionic non-local conserved quantity of
the Super Gardner equation \cite{Andrea3}, where the Gardner
category was introduced.

In \cite{Andrea2} a recursive gradient approach was proposed to
analyze the the SKdV (local) hierarchy and its local conserved
quantities. The algorithm starting from the gradient of a
conserved quantity generates, by application of operators
$P,D^{-2},K$, a new gradient of an associated new conserved
quantity. The algorithm provides all local conserved quantities as
well as the SKdV hierarchy of differential equations.

The existence of all such quantities is proven by induction using
the exact SUSY sequence introduced in \cite{Andrea2}.

One crucial step in the proof, which was missing in our previous
work, is to show that after applying $D^{-2}$ one still obtains a
local quantity which indeed is the gradient of a conserved
quantity.

In the first part of this work we present a complete proof of the
recursive gradient approach, for an initial data corresponding to
a local conserved quantity of the SKdV equation. In the second
part of this work we give a precise definition of the function
spaces where the non-local conserved quantities exist. We then
apply the recursive gradient approach to an initial data
corresponding to a fermionic non-local conserved quantity. It then
turns out that one can obtain step by step the complete structure
of fermionic non-local conserved quantities. We do not have,
however, a inductive proof as in the case of the local initial
data. Finally we introduce initial data which give rise to a new
set of non-local conserved quantities of the SKdV equations. We
find explicitly the first few of them. The new non-local conserved
quantities are bosonic, in distinction to the previously known
ones which are fermionic.

\section{The Recursive Gradient Algorithm}
The Susy KdV equation involves functions
$\mathbb{R}\rightarrow\Lambda$, with $\Lambda$ a finitely
generated exterior algebra. With $\theta\in\Lambda$ one of the
generators, the operator $D=\frac{\partial}{\partial
\theta}+\theta\frac{\partial}{\partial x}$ sends
$C^{\infty}(\mathbb{R},\Lambda)$ into itself, and interchanges the
two direct summands given by the parity of $\Lambda$.

With $C^{\infty}_\downarrow(\mathbb{R},\Lambda)$ the rapidly
diminishing functions, the formula $\frac{\partial}{\partial
\theta}\int^{\infty}_{-\infty}\Phi(x)dx$ gives a linear functional
$C^\infty_\downarrow(\mathbb{R},\Lambda)\rightarrow \Lambda$ which
vanishes on the image of $D$.

From a given $\Phi$ there arises $f(\Phi,D\Phi,D^2\Phi,\ldots)$
where $f$ can be any polynomial in several variables. Then $f$ can
give a nonlinear differential equation $\frac{\partial}{\partial
t}\Phi(x,t)=f(\Phi,D\Phi,D^2\Phi,\ldots)$, while another
polynomial $h$ might give a conserved quantity
$\frac{\partial}{\partial \theta}\int h(\Phi,D\Phi,\ldots)dx.$

In the following, an algebraic model is proposed for the study of
these questions.

The preceding scenario is replaced by a free derivation algebra on
a single fermionic generator, and the $D$ just given is replaced
by an algebraically constructed derivation designed to reflect the
general properties of the analytical $D$.

Operators, pseudodifferential operators and adjoint involutions
are described. The results are then applied to show that the
gradients of the local and non-local conserved quantities of the
supersymmetric KdV equation are generated by a recursive algorithm
formulated in this algebraic context.
\section{The derivation algebra} Let $\mathcal{A}$ be the free
supersymmetric derivation algebra on a single fermionic generator.
It is generated over the real number field by an identity element
and elements $a_1,a_2,a_3,\ldots$ subject only to the relations
$a_pa_q=a_qa_p{(-1)}^{pq}.$ Anticommutations only occur among
$a_1,a_3,a_5,\ldots$, all of whose squares are zero.

The parity involution $u\rightarrow \overline{u}$ is the algebra
automorphism of $ \mathcal{A}$ determined by $
\overline{a_p}=a_p{(-1)}^p.$

Then $ \mathcal{A}=\mathcal{A}_{even}\oplus \mathcal{A}_{odd}$ by
the $\pm 1$ eigenspaces of the parity involution, making $
\mathcal{A}$ into a supercommutative superalgebra.

The canonical superderivation $D:\mathcal{A}\rightarrow
\mathcal{A}$ will give $Da_p=a_{p+1}$ for $p\geq 1$, and satisfy
the twisted product rule \[D(uv)=(Du)v+\overline{u}(Dv)\] for all
$u,v\in \mathcal{A}.$ Furthermore $D$ reverses parity, which is to
say that $D\mathcal{A}_{odd}\subset \mathcal{A}_{even}$ and
$D\mathcal{A}_{even}\subset \mathcal{A}_{odd}$.

The Euler operator $E:\mathcal{A}\rightarrow \mathcal{A}$ will
have values $Ea_p=a_p$ for $p\geq 1$, and satisfy the ordinary
product rule \[E(uv)=(Eu)v+u(Ev),\] as well as
$E\mathcal{A}_{odd}\subset A_{odd}$ and
$E\mathcal{A}_{even}\subset A_{even}$.

The operators $D$ and $E$ are constructed from the operators
$\frac{\partial}{\partial a_p}:\mathcal{A}\rightarrow
\mathcal{A}$, as follows.

Given $1\leq p< \infty$, the complementary subalgebra $
\mathcal{A}_p\subset \mathcal{A}$ is generated by the identity
element and the $a_q$ for which $q\neq p$. Then as a vector space
direct sum \[ \mathcal{A}=\mathcal{A}_p\oplus
a_p\mathcal{A}_p\oplus a_p^2\mathcal{A}_p+\cdots\] and
$\frac{\partial}{\partial a_p}:\mathcal{A}\rightarrow \mathcal{A}$
is defined in the customary fashion.

When $p$ is even, $\frac{\partial}{\partial a_p}$ preserves parity
and satisfies the ordinary product rule. When $p$ is odd the
direct sum reduces to $ \mathcal{A}_p\oplus a_p\mathcal{A}_p$, and
$\frac{\partial}{\partial a_p}$ reverses parity and satisfies the
twisted product rule. These operators satisfy the commutation rule
$ \frac{\partial}{\partial a_p}\frac{\partial}{\partial
a_q}=\frac{\partial}{\partial a_q}\frac{\partial}{\partial
a_p}{(-1)}^{pq}.$

The claimed properties of $D$ and $E$ then follow from the
explicit formulas
\[D=a_2\frac{\partial}{\partial a_1}+a_3\frac{\partial}{\partial
a_2}+\cdots \]
\[E=a_1\frac{\partial}{\partial a_1}+a_2\frac{\partial}{\partial
a_2}+\cdots\] Furthermore the commutator
$(DE-ED):\mathcal{A}\rightarrow \mathcal{A}$ satisfies the twisted
product rule; the operator identity $DE=ED$ then follows from its
truth on the generating elements $a_1,a_2,\ldots$ This shows that
$D$ preserves the homogeneous subspaces of $ \mathcal{A}$, that
is, the eigenspaces of the Euler operator $E$.

Given $u\in \mathcal{A}$, the possibility $u\in D\mathcal{A}$ is
now investigated. Using the congruence notation $u\equiv v$ when
$u-v\in D\mathcal{A}$, the general fact $(D^2f)g\equiv -f(D^2g)$
when applied to \[Eu=a_1\frac{\partial u}{\partial
a_1}+a_2\frac{\partial u}{\partial a_2}+\cdots\] gives
\begin{eqnarray*}Eu &\equiv & a_1\left(\frac{\partial u}{\partial
a_1}-D^2\frac{\partial u}{\partial a_3}+
D^4\frac{\partial u}{\partial a_5}-\cdots \right) \\
&+& a_2\left(\frac{\partial u}{\partial a_2}-D^2\frac{\partial
u}{\partial a_4}+ D^4\frac{\partial u}{\partial a_6}-\cdots
\right).\end{eqnarray*} Another general fact $a_2h\equiv a_1Dh$
then gives \[Eu\equiv a_1Mu\] where $M:\mathcal{A}\rightarrow
\mathcal{A}$ is the gradient operator \[M=\frac{\partial}{\partial
a_1}+D\frac{\partial}{\partial a_2}-D^2\frac{\partial}{\partial
a_3}-D^3\frac{\partial}{\partial a_4}+\cdots\] Evidently the
condition $Mu=0$ implies $Eu\in D\mathcal{A}$; if $u$ has zero
constant term its homogeneous components and hence $u$ itself are
in $D\mathcal{A}.$ It is also true that $MD\equiv 0$ as an
operator $\mathcal{A}\rightarrow \mathcal{A}$.
\section{The algebras $\mathcal{O}_p\mathcal{A}\subset
\mathcal{P}_{sd}\mathcal{A}$} When the operator $D$ acts on the
product of two elements of $\mathcal{A}$, the result is
\[D(uv)=\overline{u}Dv+(Du)v.\] For higher powers of $D$, the
supersymmetric binomial coefficients are needed. They are given by
the generating functions \[F_m(x)=\sum_{p=0}^\infty
\left[\begin{array}{c}m\\p \end{array}\right]x^p\] in which
\[F_m(x)=\left\{\begin{array}{l}{(1+x^2)}^n \hspace{3mm}\mathrm{\:when\:}m=2n\\
{(1+x^2)}^n(1+x)\hspace{3mm}\mathrm{\:when\:}m=2n+1.\end{array}\right.
\] For $\overline{u}=\pm u$ one can use the notation
$u=u_M,\overline{u_M}=u_M{(-1)}^M.$ The images of $u_M$ under
repeated applications of $D$ can then be written $D^pu_M=u_{M+p}$
with $ \overline{u}_{M+p}=u_{M+p}{(-1)}^{M+p}.$

Then for any integer $m>0$ the appropiate Leibnitz formula is
\[D^m(u_Mv)=\sum_{p=0}^m{(-1)}^{M(m+p)}\left[\begin{array}{c}m\\p \end{array}\right]u_{M+p}D^{m-p}v.\]
For $m=1$ it gives $D(u_Mv)={(-1)}^Mu_MDv+u_{M+1}v$ as it should.
For a proof by induction one passes from $m$ to $m+1$ by computing
\[D(u_{M+p}D^{m-p}v)={(-1)}^{M+p}u_{M+p}D^{m+1-p}v+u_{M+p+1}D^{m-p}v.\]
The identity \[\left[\begin{array}{c}m\\p-1
\end{array}\right]+{(-1)}^p\left[\begin{array}{c}m\\p
\end{array}\right]=\left[\begin{array}{c}m+1\\p \end{array}\right]
\] then gives the desired result: it follows from the recursion
\[xF_m(x)+F_m(-x)=F_{m+1}(x)\] satisfied by the generating
functions.

The algebra $ \mathcal{O}_p\mathcal{A}$ consists of the linear
transformations $L:\mathcal{A}\rightarrow \mathcal{A}$ which have
the form $L=\sum_0^Nl_nD^n$ for some $l_n\in \mathcal{A}$ and
$0\leq N<\infty$. The parity involution of $
\mathcal{O}_p\mathcal{A}$ sends $L$ to $
\overline{L}=\sum_0^N\overline{l_n}{(-D)}^n.$

Thus $L$ is ``oriented" when $ \overline{L}=\pm L$, and
\[\mathcal{O}_p\mathcal{A}={(\mathcal{O}_p\mathcal{A})}_{even}\oplus
{(\mathcal{O}_p\mathcal{A})}_{odd}\]is a superalgebra.

The product of oriented operators is defined by bilinear expansion
from the special case
\[\left(u_MD^m \right)\left(v_ND^n\right)=\sum_{p\geq 0}{(-1)}^{N(m+p)}\left[\begin{array}{c}m\\p \end{array}\right]
u_Mv_{N+p}D^{m+n-p}.\] Since the product of operators is defined
independently as the composition of linear transformations of a
vector space, the associativity of the product would seem to be
clear.

But there are no negative powers of $D$ in $
\mathcal{O}_p\mathcal{A}.$ For this reason $
\mathcal{O}_p\mathcal{A}$ is enlarged to $P_{sd}\mathcal{A},$
whose elements are the formal semi-infinite sums
\[L=\sum_{-\infty}^Nl_nD^n\] with $l_n\in \mathcal{A}$ and
$-\infty<N<\infty$. The same parity involution is present, and the
product of two oriented elements of $ \mathcal{P}_{sd}\mathcal{A}$
is given by bilinear expansion using the same formula for
$(u_MD^m)(v_ND^n)$ as in $\mathcal{O}_p\mathcal{A}$, but with
$0\leq p<\infty$.

When $m<0$ the coefficients $\left[\begin{array}{c}m\\p
\end{array}\right]$ do not vanish identically for $p>>0$, and they
leave the product as a semi-infinite formal sum.

The associativity equation $A(BC)=(AB)C$ must now be established
for any three elements $A,B,C\in \mathcal{P}_{sd}\mathcal{A}.$

A first observation is that any equation $A(BC)=(AB)C$ in
$\mathcal{P}_{sd}\mathcal{A}$ may be multiplied on the left by
$h_kI$ and on the right by $D^r$, giving another such equation
\[ (h_kA)\left(B(CD^r)  \right)=\left((h_kA)B  \right)(CD^r).  \]
Then two more special cases are sufficient for the general result.
First, when\[D^m(u_MIv_NI)=(D^m(u_MI))(v_NI) \] is expanded, it is
seen to follow from the cancellation identity
\[\left[\begin{array}{c}m\\p \end{array}\right]\left[\begin{array}{c}m-p\\q
\end{array}\right]=
\left[\begin{array}{c}m\\p+q
\end{array}\right]\left[\begin{array}{c}p+q\\p
\end{array}\right]\] which holds for all $p,q\geq0$ and
$-\infty<m<\infty$.

Second, when \[D^m\left(D^n\left(v_NI \right)
\right)=D^{n+m}\left(v_NI\right)
\] is worked out, it is seen to follow from the
``sum-of-exponents" identities
\[\left[\begin{array}{c}n+m\\p
\end{array}\right]=\sum_{\begin{array}{l}r+s=p \\ r\geq0 \\s\geq0
\end{array}}\left[\begin{array}{c}n\\r \end{array}\right]
\left[\begin{array}{c}m\\s \end{array}\right]{(-1)}^{r(m+p+1)}.\]
 These identities in turn follow
from the equations connecting the generating function $F_{n+m}(x)$
with $F_n(\pm x)$ and $F_m(x).$

Multiplying on the left by elements of $\mathcal{A}$ and on the
right by powers of $D$, we obtain general elements of
$\mathcal{P}_{sd}\mathcal{A}.$

Thus $A(BC)=(AB)C$ is proven when $B=h_kI$ or $B=D^r$.

Finally, to prove associativity for three elements $D^m,u_MD^n,$
and $v_NI$, we compute \begin{eqnarray*}D^m\left(\left(u_MD^n
\right)\left(v_NI \right) \right)
&=&D^m\left(u_M\left(D^n\left(v_NI \right) \right) \right)\\
&=&\left(D^m\left(u_MI \right) \right)\left( D^n\left(v_NI
\right)\right)\end{eqnarray*} and
\begin{eqnarray*}\left(D^m\left(u_MD^n
\right)\right)v_NI   &=& \left(\left(D^m\left(u_MI \right)
\right)D^n \right)\left(v_NI \right) \\ &=& \left(D^m\left(u_MI
\right) \right)\left({D^n}\left(v_NI\right) \right).
\end{eqnarray*} This equality completes the proof that
$\mathcal{P}_{sd}\mathcal{A}$ is an associative algebra.

A sample formula in $\mathcal{P}_{sd}\mathcal{A}$ is
\[D^{-2}(uI)=uD^{-2}-(D^2u)D^{-4}+(D^4u)D^{-6}-\cdots\in
\mathcal{P}_{sd}\mathcal{A};\] it will be used in the applications
which follow.
\section{The Adjoint Involution}
The parity-preserving involution $L\rightarrow L^*$ of
$\mathcal{P}_{sd}\mathcal{A}$ with itself is determined by the
three properties
\begin{eqnarray*}\hspace{3mm}& & (1)\hspace{1mm} {(uI)}^*=uI \hspace{2mm} \mathrm{\:for\:} u\in\mathcal{A}
\\& & (2)\hspace{1mm} D^*=-D \\ & &  (3)\hspace{1mm} {(L_1L_2)}^*={(-1)}^{\lambda_1\lambda_2}L_2^*L_1^*, \end{eqnarray*}
when $L_1,L_2\in\mathcal{P}_{sd}\mathcal{A}$ have parities
${(-1)}^{\lambda_1}$ and ${(-1)}^{\lambda_2}.$

The last two properties when applied to powers of $D$ give
\[{(D^n)}^*={(-1)}^{\frac{n(n+1)}{2}}D^n.
\] Then, when $u_N\in\mathcal{A}$ has parity ${(-1)}^N$, the
adjoint of $L=u_ND^n$ must be defined by
\begin{eqnarray*}L^* &=& {(-1)}^{nN}{(D^n)}^*{(u_NI)}\\ &=& {(-1)}^{\frac{n(n+1)}{2}}
\sum_{p=0}^\infty{(-1)}^{Np}\left[\begin{array}{c}n\\p
\end{array}\right]u_{N+p}D^{n-p}  \end{eqnarray*} with
$u_{N+p}=D^pu_N$ as before. Since every element of
$\mathcal{P}_{sd}\mathcal{A}$ is a formal sum of powers of $D$
multiplied from the left by elements of $\mathcal{A}$, this
construction gives a well-defined linear transformation
$L\rightarrow L^*$ of $\mathcal{P}_{sd}\mathcal{A}$ into itself.

But to verify (3) when $L_1=D^m$ and $L_2=u_NI,$ the product
$D^m(u_NI)$ must first be expanded as a infinite linear
combination of $u_{N+q}D^{m-q}$ with $q\geq0$, and then the
$L\rightarrow L^*$ construction just given must be applied to each
term. The result is a double summation over $q\geq0,p\geq0$, and
property (3) reduces to the identities
\[\sum_{p+q=r>0}{(-1 )}^{pq+\frac{1}{2}(m-q)(m+1-q)
}\left[\begin{array}{c}m\\q
\end{array}\right]\left[\begin{array}{c}m-q\\p
\end{array}\right]=0.   \] The cancellation identity of the last
section puts this into the form \[\left[\begin{array}{c}m\\r
\end{array}\right]\sum_{q=0}^r\varepsilon(q)\left[\begin{array}{c}r\\q
\end{array}\right]=0  \] with
$\varepsilon(q)={(-1)}^{q(r-q)+\frac{1}{2}(m-q)(m+1-q)}.$

When $r\geq2$ the generating function $F_r(x)$ satisfies
$F_r(i)=0$ with $i^2=-1$, giving
\begin{eqnarray*}0&=&\left[\begin{array}{c}r\\0
\end{array}\right]-\left[\begin{array}{c}r\\2
\end{array}\right]+\left[\begin{array}{c}r\\4
\end{array}\right]-\cdots \\0&=&\left[\begin{array}{c}r\\1
\end{array}\right]-\left[\begin{array}{c}r\\3\end{array}\right]+\left[\begin{array}{c}r\\5
\end{array}\right]-\cdots
\end{eqnarray*} Since $\varepsilon(q+2)=-\varepsilon(q)$ for all
integers $q$, the identity is proved in the case $r\geq2$.

In the remaining case $r=1$, $\left[\begin{array}{c}m\\r
\end{array}\right]   $ can be nonzero only when $m$ is odd: then
$\varepsilon(0)+\varepsilon(1)=0$ and $\left[\begin{array}{c}r\\0
\end{array}\right]=\left[\begin{array}{c}r\\1
\end{array}\right]=1.   $

Thus (3) is confirmed in all the four cases where $L_1$ and $L_2$
can be $u_NI,v_MI,$ or powers of $D$. Using associativity in
$v_Mu_ND^n$ and $u_ND^nD^m$, these four cases give
\begin{eqnarray*}{(v_ML )}^* &=& {(-1)}^{M\lambda }L^*{(v_MI)}^* \\ {(LD^m)}^* &=&
{(-1)}^{\lambda m}{(D^m)}^*L^* \end{eqnarray*} for $L=u_ND^n$ and
$\lambda=N+n.$

Finally, the general case $L_1L_2=u_ND^nv_MD^m$ can be expanded by
applying the preceding special cases. This gives
\begin{eqnarray*}{(L_1L_2)}^*&=&\pm
{(D^m)}^*{(v_MI)}^*{(D^n)}^*{(u_NI)}^* \\ &=& \pm
L_2^*L_1^*,\end{eqnarray*} the $\pm$ sign given by
${(-1)}^{\lambda_1\lambda_2}$ with $\lambda_1=N+n,\lambda_2=M+m.$

This completes the proof of (3) for all elements of
$\mathcal{P}_{sd}\mathcal{A}$. The involutive property
${(L^*)}^*=L$ is a direct consequence.

\section{The Frechet derivative operator} The construction
$h\rightarrow L_h$ which takes $h\in\mathcal{A}$ to its Frechet
derivative operator $L_h\in\mathcal{O}_p\mathcal{A}$ is now
described.

For odd elements $f=-\overline{f}$ in $\mathcal{A}$ the action of
$L_h$ is given by
\begin{eqnarray*}h &=& h(a_1,a_2,a_3,\ldots) \\ L_hf &=&
\frac{d}{dt}\big{|}^{t=0}h(a_1+tf,a_2+tDf,a_3+tD^2f,\ldots).
\end{eqnarray*}
For fixed $f$ and varying $h$, the transformation $
\mathcal{A}\rightarrow\mathcal{A}$ given by $Fh= L_hf$ preserves
parity and satisfies the ordinary product rule
$F(h_1h_2)=(Fh_1)h_2+h_1(Fh_2).$ Further, $Fa_p= D^{p-1}f$ for all
$p\geq1$.

On the other hand $D:\mathcal{A}\rightarrow\mathcal{A}$ reverses
parity and satisfies the twisted product rule. With $Fh=L_hf$, the
commutator $(DF-FD):\mathcal{A}\rightarrow\mathcal{A}$ satisfies
the twisted product rule and gives the value zero on all the
generators $a_p$. This shows that $\left[ D,F\right]=0$ on all
elements of $\mathcal{A}$, proving that \[DL_hf=L_{Dh}f. \] A
second consequence is the explicit formula
\[L_hf=f\frac{\partial h}{\partial a_1}+(Df)\frac{\partial h}{\partial a_2}
+(D^2f)\frac{\partial h}{\partial a_3}+\cdots \] Indeed, this
formula sends $h=a_p$ to $Fh=D^{p-1}f$ and satisfies the ordinary
product rule when applied to $h_1h_2$: therefore it must coincide
with the $L_hf$ given by the definition not just on the generators
$a_p$ but everywhere in $\mathcal{A}.$

The explicit formula gives $L_ha_1=Eh$ and $L_ha_3=D^2h,$ for
example. And when $h$ is oriented with $
\overline{h}=h{(-1)}^\chi,$ reorderings and sign changes put
$L_hf$ into the standard form $\sum l_nD^nf.$

For oriented elements $h$, the construction $h\rightarrow L_h$
reverses parity, in the sense that when $h$ has parity
${(-1)}^\chi$, $L_h$ has parity ${(-1)}^{(\chi+1)}.$When written
out explicitly the Frechet derivative operator is
\[L_h=\sum_{n=1}^\infty {(-1)}^{n(\chi+1)}\frac{\partial h}{\partial a_n}D^{n-1}, \]
giving $L_g=-\left(\frac{\partial g}{\partial
a_1}\right)I+\left(\frac{\partial g}{\partial a_2}
\right)D-\left(\frac{\partial g}{\partial a_3}\right)D^2+\cdots$
for example when $ \overline{g}=g$ and $\chi$.

When taken together with the construction of adjoint operators,
there are two important applications of the Frechet derivative.
The first is an analog of the mixed partials criterion in the
Poincare lemma:

if $g\in \mathcal{A}$ satisfies $L_g+L_g^*=0$ then $g$ is the
gradient $Mh$ of some $h\in \mathcal{A},$ by the exact sequence of
calculus of variations.

The second application characterizes those $h\in\mathcal{A}$ which
fall into $D^2\mathcal{A}\subset\mathcal{A}.$

Indeed, $h=D^2l$ implies $L_h=D^2L_l,$ which says that $L_h=D^2Q$
for some $Q\in \mathcal{O}_p\mathcal{A}$. Conversely, applying
$L_h=D^2Q$ to the generating element $a_1$, we obtain $Eh\in
D^2\mathcal{A}.$ Since $D$ and $E$ commute, the equation $Eh=D^2u$
resolves into homogeneous components, giving
\[ h=D^2\left(u_1+\frac{1}{2}u_2+\frac{1}{3}u_3+\cdots\right)\in D^2\mathcal{A}. \]

Taking adjoints, $L_h=D^2Q$ becomes $L_h^*=-Q^*D^2.$ This means
that $h\in \mathcal{A}$ with zero constant term will fall into
$D^2\mathcal{A}$ if and only if the bottom two coefficients of the
adjoint of its Frechet derivative operator are zero, that is,
\[L_h^*=0\cdot I+0\cdot D+(?)D^2+\cdots\]

The same reasoning when applied to $D$ instead of $D^2$ suggests
that \[ L_h^*=(\pm Mh)I+(?)D+\cdots,\] this is indeed the case
when $h=\pm\overline{h}.$
\section{The Recursion} The recursive algorithm for the gradients
of conserved quantities claims the existence of even elements
$g_2,g_4,g_6,\ldots$ and odd elements $f_3,f_5,f_7,\ldots$ in
$\mathcal{A}$ which satisfy \begin{eqnarray*}& & Pg_n=D^2f_{n+1} \\
& & Kf_{n+1}=D^2g_{n+2}
\end{eqnarray*} for the operators
\begin{eqnarray*}P&=&D^5+3a_1D^2+a_2D+2a_3I \\K&=&D^3+a_1I.
\end{eqnarray*} The additional condition $L_{g_n}+L_{g_n}^*=0$ is,
by the exact sequence, equivalent to the existence of
$h_n\in\mathcal{A}$ having $g_n$ as its gradient.

By direct computation one can check that the choice
$g_2=a_2,f_3=a_5+3a_1a_2,g_4=a_6+3a_2^2-2a_1a_3$ satisfies the
recursion, and that the Frechet derivative operators $L_{g_2}=D$
and $L_{g_4}=D^5-2a_1D^2+6a_2D+2a_3I$ are antisymmetric. It then
remains to be shown that the algorithm continues indefinitely.

The operators $P$ and $K$ appearing in the recursion are both odd,
with adjoints $K^*=K$ and $P^*=-P.$ When $g=\overline{g}$ and $
\overline{f}=-f$ in $\mathcal{A}$, the Frechet derivative
operators of $Pg$ and $Kf$ are given by
\begin{eqnarray*}& & L_{Kf}=KL_f-fI \\ & & L_{Pg}=PL_g+R_g \\ & & R_g=2gD^2+(Dg)D+(3D^2g)I  \end{eqnarray*}
and their adjoints by
\begin{eqnarray*}& & L_{Kf}^*=L_f^*K-fI \\& & L_{Pg}^*=L_g^*P+R_g^* \\& & R_g^*=-2gD^2-(Dg)D+
(2D^2g)I.  \end{eqnarray*}

Given a satisfactory choice of $g_2,f_3$ and $g_4$, it must now be
shown that $f_5$ exists, that $g_6$ exists, and that $Lg_6$ is
antisymmetric.

The existence of $f_5\in\mathcal{A}$ with $D^2f_5=Pg_4$ is
determined by the Frechet derivative operator $L_{Pg_4}$ whose
adjoint $L_{Pg_4}^*$ must be shown to have bottom two coefficients
zero.

Using $L_{g_4}^*=-L_{g_4}$ this means that
\[L_{g_4}P-R_{g_4}^*=0\cdot I+0\cdot D+\cdots,
\] which in turn would follow from the general fact
\[ L_gP=(2D^2g)I-(Dg)D+\cdots \] for any $ \overline{g}=g$ in
$\mathcal{A}$.

If we compute \begin{eqnarray*}& & P=2a_3I+a_2D+\cdots \\ & &
DP=2a_4I-a_3D+\cdots \\ & & D^2P=2a_5I+a_4D+\cdots
\end{eqnarray*} and recall that \[L_g=-\frac{\partial g}{\partial
a_1}I+\frac{\partial g}{\partial a_2}D-\frac{\partial g}{\partial
a_3}D^2+\cdots\] we obtain \[L_gP=l_0I+l_1D+\cdots \] with
\begin{eqnarray*}& & l_0=2\left(-\frac{\partial g}{\partial
a_1}a_3+\frac{\partial g}{\partial a_2}a_4-\frac{\partial
g}{\partial a_3}a_5+\cdots  \right) \\ & & l_1=
-\left(\frac{\partial g}{\partial a_1}a_2+\frac{\partial
g}{\partial a_2}a_3+\frac{\partial g}{\partial a_3}a_4+\cdots
\right).
\end{eqnarray*}Because $g$ is even, $\frac{\partial g}{\partial
a_p}a_q=a_q\frac{\partial g}{\partial a_p}{(-1)}^{pq}.$

This, together with $D^2=a_3\frac{\partial}{\partial
a_1}+a_4\frac{\partial}{\partial a_2}+\cdots,$ proves that
$l_0=2D^2g$ and $l_1=-Dg.$

With the general fact established, the existence of $f_5$ is
proven.

The existence of $g_6$ with $D^2g_6=Kf_5$ is determined by the
Frechet derivative operator $L_{Kf_5}=KL_{f_5}-f_5I,$ whose
adjoint $L_{f_5}^*K-f_5I$ must be shown to have bottom two
coefficients zero.

Since $D^2(l_0I+l_1D^2+\cdots)=(D^2l_0)I+(D^2l_1)D+\cdots$ and
$D:\mathcal{A}\rightarrow \mathcal{A}$ is injective, it suffices
to prove that
\[ D^2L_{f_5}^*K=(Pg_4)I+0\cdot D+\cdots   \]

But because the adjoint of this operator has bottom two
coefficients zero,it is enough to examine
\begin{eqnarray*}& & G=KL_{f_5}D^2 \\ & & G^*=-D^2L_{f_5}^*K,
\end{eqnarray*}and to prove that \[ G+G^*=-(Pg_4)I+0\cdot D+\cdots\]
The equation $D^2f_5=Pg_4$ gives $D^2L_{f_5}=PL_{g_4}+R_{g_4}$ and
hence
\[G=KD^{-2}R_{g_4}D^2+KD^{-2}PL_{g_4}D^2.\]
Then, the recursions $D^2g_4=Kf_3$ and $D^2f_3=Pg_2$ permit
$L_{f_3}$ to be eliminated between the equations
$D^2L_{g_4}=KL_{f_3}-f_3I$ and $D^2L_{f_3}=PL_{g_2}+R_{g_2},$
giving \[D^2L_{g_4}=-f_3I+KD^{-2}R_{g_2}+KD^{-2}PL_{g_2}. \]
Taking the adjoint of this equation and remembering the
antisymmetry of $L_{g_4}$ and $L_{g_2}$, we get
\[L_{g_4}D^2=-f_3I-R_{g_2}^*D^{-2}K+L_{g_2}PD^{-2}K.\]
This equation permits $G$ to be rewritten as
\begin{eqnarray*}G= & & KD^{-2}R_{g_4}D^2 -KD^{-2}P(f_3I) \\& &
-KD^{-2}PR_{g_2}^*D^{-2}K+KD^{-2}PL_{g_2}PD^{-2}K. \end{eqnarray*}
What must then be shown is that $G+G^*=-(Pg_4)I+0\cdot D+\cdots;$
from the original definition
$G=KL_{f_5}D^2\in\mathcal{O}_p\mathcal{A}$ it is clear that no
negative powers of $D$ will enter.(In fact it turns out that
$G+G^*=(-Pg_4)I$ exactly.)

Of the four summands in $G$, all but one are elements of
$\mathcal{P}_{sd}\mathcal{A}$ of order $\leq 9.$ The four summands
and their adjoints are \[\begin{array}{ll} A=KD^{-2}R_{g_4}D^2 &
A^*=D^2R_{g_4}^*D^{-2}K \\B=-KD^{-2}P(f_3I) & B^*=f_3PD^{-2}K \\
C=-KD^{-2}PR_{g_2}^*D^{-2}K & C^*=-KD^{-2}R_{g_2}PD^{-2}K \\
F=KD^{-2}PL_{g_2}PD^{-2}K & F^*=KD^{-2}PL_{g_2}^*PD^{-2}K.
\end{array}\] Of these operators only $F$ might have order $> 9$.
However of the seven factors appearing in $F$ five are odd, and
${5\choose 2}=10$ an even number. Therefore $F^*$ is as stated
above without a minus sign, and the induction hypothesis of the
antisymmetry of $L_{g_2}$ gives $F+F^*=0.$

Regarding $C$, it can be proved that
$PR_g^*+R_gP=(3Pg)D^2+(DPg)D+(2D^2Pg)I$ for any $g=\overline{g}$
in $\mathcal{A}$. Since $Pg_2=D^2f_3$ we have
\[C+C^*=-KD^{-2}LD^{-2}K \] with
$L=(3D^2f_3)D^2+(D^3f_3)D+(2D^4f_3)I.$

The coefficients of $A+A^*,B+B^*,C+C^*$ for nonnegative powers of
$D$ can be computed. Summing them to get the coefficients of
$G+G^*$,we begin with the positive powers and write
\[\begin{array}{ccc} \hspace{20mm} \underline{A+A^*} &
\underline{B+B^*} & \underline{C+C^*} \\{(G+G^*)}_1=2D^4g_4 &
-2a_1D^2f_3-2a_3f_3 & -2D^5f_3 \\ {(G+G^*)}_2=2D^3g_4 &
-3D^4f_3+2a_1Df_3-2a_2f_3 & D^4f_3 \\ {(G+G^*)}_3=4D^2g_4 &
-4a_1f_3 & -4D^3f_3 \\ {(G+G^*)}_4=0 & -3D^2f_3 & 3D^2f_3 \\
{(G+G^*)}_5=0 & 0 & 0
\end{array}
\]
The recursion $D^2g_4=D^3f_3+a_1f_3$ shows that ${(G+G^*)}_3=0;$
likewise $D^3g_4=D^4f_3-a_1Df_3+a_2f_3$ and
$D^4g_4=D^5f_3+a_1D^2f_3+a_3f_3$ proving that ${(G+G^*)}_n=0$ for
all $n\geq 1$. It only remains to compute ${(G+G^*)}_0$, the
coefficient of the identity operator.

We have
\begin{eqnarray*}{(G+G^*)}_0= & & (a_1D^2-a_2D-2a_3)g_4 \\ & & +(-D^6+2a_1D^3-
4a_2D^2+a_3D-a_4I)f_3 \\ & & +(-5a_1D^3+3a_2D^2)f_3.
\end{eqnarray*}

Adding \[Pg_4=(D^5+3a_1D^2+a_2D+2a_3I)g_4 \] we obtain
\[{(G+G^*)}_0+Pg_4=(D^5+4a_1D^2)g_4+(-D^6-3a_1D^3-a_2D^2+a_3D-a_4)f_3.\]

Since $D^2g_4=D^3f_3+a_1f_3$, and since the product of operators
gives
\begin{eqnarray*}(D^3+4a_1I)(D^3+a_1I)\\ &=&D^6+3a_1D^3+a_2D^2-a_3D+a_4I,
\end{eqnarray*}
we have ${(G+G^*)}_0+Pg_4=0.$

This completes the proof that $G+G^*=-(Pg_4)I$ when
$G=KL_{f_5}D^2,$ and consequently that $g_6\in \mathcal{A}_{even}$
exists with $D^2g_6=Kf_5.$

It only remains to carry out the third and final step, which is to
prove $L_{g_6}+L_{g_6}^*=0.$

This will follow from the equations
\begin{eqnarray*}D^2L_{g_6}&=&KL_{f_5}-f_5I \\ L_{g_6}^*D^2 &=& -L_{f_5}^*K+f_5I \end{eqnarray*}
which give
\begin{eqnarray*}D^2(L_{g_6}+L_{g_6}^*)D^2 &=& G+G^*-f_5D^2 +D^2(f_5I) \\ &=&
(-Pg_4)I+(D^2f_5)I \\&=&0.   \end{eqnarray*} Since
$L_{g_6}+L_{g_6}^*\in\mathcal{O}_p\mathcal{A}$ with $D^2$ an
invertible element of $\mathcal{P}_{sd}\mathcal{A},$ the result
follows.

This completes the proof of the indefinite continuation of the
recursive algorithm for the gradients of the local conserved
quantities of the Susy KdV equation.
\section{Rings of Superfields} A superfield is an infinitely differentiable
function $\Phi:\mathbb{R}\rightarrow\Lambda$. The ring
$C_{NL}^\infty$ consists of the nonlocal superfields, those that
diminish rapidly at $x=-\infty$ and increase slowly at
$x=+\infty$. This means that $\lim {|x|}^N\Phi(x)=0$ for all
$-\infty<N<\infty$ when $x\rightarrow-\infty$ and for some $N$ as
$x\rightarrow +\infty$. The same condition is assumed to hold for
all $({\frac{\partial}{\partial x}})^p\Phi(x).$

Two more rings of superfields satisfy the inclusions
\[C_\downarrow^\infty\subset C_I^\infty\subset C_{NL}^\infty.   \]
The ideal $C_\downarrow^\infty$ is the Schwartz space of
superfields which diminish rapidly at $\pm\infty$ together with
all $x$-derivatives while $C_I^\infty$ is defined by the condition
$\frac{\partial}{\partial \theta}\Phi\in C_\downarrow^\infty.$

The rings $C_\downarrow^\infty$ and $C_{NL}^\infty$ are invariant
under the action of the operator $D$. In the smaller ring $D$ is
not invertible, $DC_\downarrow^\infty$ being only a proper
subspace. But in the larger ring $C_{NL}^\infty$ the formulas
\[\Phi(x)=\xi(x)+\theta u(x)   \]
\[(D^{-1}\Phi)(x)=\int_{-\infty}^xu(s)ds+\theta \xi(x)\]
give $D^{-1}:C_{NL}^\infty\rightarrow C_{NL}^\infty,$ the inverse
to the bijection of $C_{NL}^\infty$ with itself which is given by
the operator $D$.
\section{Integration} The integration linear functional
$C_I^\infty\rightarrow\Lambda$ is given by
\[\Phi(x)=\xi(x)+\theta u(x)\]
\[\int\Phi=\int_{-\infty}^\infty u(x)dx.\]
Evidently $\int D\Phi=0$ when $\Phi\in C_\downarrow^\infty.$

Because $C_\downarrow^\infty$ is an ideal in $C_{NL}^\infty$, one
has integration-by-parts formulas
\[\int(D\Phi)\Psi=\pm\int\Phi(D\Psi)\]
when $\Psi\in C_{NL}^\infty$ and $\Phi\in C_\downarrow^\infty$ is
oriented.

\section{Gradients} A function
$H:C_\downarrow^\infty\rightarrow\Lambda$ may be said to have
another function $\Gamma:C_\downarrow^\infty\rightarrow
C_{NL}^\infty$ as its gradient if, for any $\Phi,\dot{\Phi}\in
C_\downarrow^\infty,$
\[ \frac{d}{dt}|^{t=0}H(\Phi+t\dot{\Phi})=\int\dot{\Phi}\Gamma(\Phi).  \]
In what follows $H$ will have the form $H(\Phi)=\int h(\Phi)$ for
some $h:C_\downarrow^\infty\rightarrow C_I^\infty.$

To know that $H$ is a conserved quantity for a differential
equation, the preceding equation need only hold when $ \dot{\Phi}$
is given in terms of $\Phi,D\Phi,\ldots$ by the differential
equation, provided that $\int\dot{\Phi}\Gamma(\Phi)=0.$ Then
$\Gamma$ can be called a ``restricted" gradient of $H$.

\section{The Recursive Algorithm} Given an odd $\Phi\in
C_\downarrow^\infty$, two operators acting on superfields are
given by \begin{eqnarray*}& & P=D^5+3\Phi D^2+(D\Phi)D+(2D^2\Phi)I \\
& & K=D^3+\Phi I.  \end{eqnarray*} Then five superfields
$\Gamma_0,\Omega_1,\Gamma_2,\Omega_3,\Gamma_4$ satisfy the
recursion if
\begin{eqnarray*} & & P\Gamma_0=\Omega_1=D^2\Gamma_2 \\ & &
K\Gamma_2=\Omega_3=D^2\Gamma_4.\end{eqnarray*}

An infinite sequence of superfields
$\left\{,\ldots,\Omega_{-1},\Gamma_0,\Omega_1,\Gamma_2,\ldots,
\right\}$ satisfies the recursion if
$\Gamma_m,\Omega_{m+1},\ldots,\Gamma_{m+4}$ are connected by the
same equations when $m=0,\pm 4,\pm 8,\ldots$

Supposing $\Omega_n=\Gamma_m=0$ for negative integers, the choice
of initial value $\Gamma_0=\frac{1}{2}$ has been shown in the
preceding sections to produce $\Omega_1,\Gamma_2,\ldots$ that stay
within $C_\downarrow^\infty$, despite the apparent presence of
$D^{-2}$ in the recursion.

Moreover, $\Gamma_0=\frac{1}{2}$ gives
\[ \Omega_5=D^6\Phi+3\Phi D^3\Phi+3(D\Phi)(D^2\Phi),  \]
which defines the SUSY K-dV equation \[\frac{\partial}{\partial
t}\Phi(x,t)=\Omega_5(\Phi,D\Phi,\ldots)
\] for time-dependent odd superfields in $C_\downarrow^\infty$.

\section{Local Conserved Quantities} It was also proved that
the superfields $\Gamma_4,\Gamma_8,\ldots\in C_\downarrow^\infty$
produced by $\Gamma_0=\frac{1}{2}$ are all gradients of functions
$H_m(\Phi)=\int h_m:C_\downarrow^\infty\rightarrow \Lambda$ where
the $h_m$ are again polynomials in $\Phi,D\Phi,\ldots,$ according
to the SUSY exact sequence.

The proof that the $H_m$ are conserved quantities follows from the
operators that appear in the recursion:

If $L$ is any one of the three operators $P,K$ and $D^2$ one has
\[\int (Lf)g=\pm\int f(Lg)   \] when $f$ and $g$ are oriented
elements of $C_{NL}^\infty$, and at least one of them is in
$C_\downarrow^\infty$.

Since $\Omega_n\in C_\downarrow^\infty$ for all odd $n$, this
gives
\[\int \Omega_n\Gamma_m=\int \Omega_{n-4}\Gamma_{m+4}=\cdots 0  \] for
all odd $n$ and even $m$, in consequence of the recursion
relations.

This shows that the $H_m$ are conserved quantities for the
differential equations given by the $\Omega_n.$

\section{Nonlocal Conserved Quantities} In general the sucessive
aplication of the operators $P,K$ and $D^{-2}$ can only be
expected to produce superfields in $C_{NL}^\infty.$

Nonetheless, other choices of initial values such as
$\Gamma_0=\theta,\Gamma_2=1,\Gamma_2=\theta$ will produce infinite
sequences of superfields satisfying the recursion, because
$D^2\theta=D^21=0.$

The choice $\Gamma_0=\frac{1}{2}$ produces
$\Omega_1,\Gamma_2,\ldots$ that stay within $C_\downarrow^\infty.$
The other three choices produce superfields in $C_{NL}^\infty$.
The initial values $\Gamma_0=\theta,\Gamma_2=1$ produce the
infinite sequence of gradients and non-local fermionic conserved
quantities already known in the literature \cite{Dargis, Andrea1}.
The initial value $\Gamma_2=\theta$ give rise to new non-local
bosonic conserved quantities.

If
$\left\{,\ldots,\widetilde{\Gamma}_m,\widetilde{\Omega}_{m+1},\ldots
\right\}\subset C_{NL}^\infty$ is such a sequence then
$\Omega_n\widetilde{\Gamma}_m=0$ continues to hold for even $m$
and odd $n$ if $\widetilde{\Gamma}_m=\widetilde{\Omega}_{m+1}=0$
when $m<<0$. This suggests that some $
\widetilde{H}_m:C_\downarrow^\infty\rightarrow\Lambda$ may exist
having $\widetilde{\Gamma_m}$ as its gradient.

If so, $\widetilde{\Gamma_m}$ would be a nonlocal conserved
quantity for the SUSY KdV equation.

This possibility is checked for the initial value
$\widetilde{\Gamma}_2=\theta$, and is seen to hold at least for
the first two gradients. The computations follow.

\section{New Non-local Conserved Quantities}
 Writing $a_1=\Phi$, any odd element of
$C_\downarrow^\infty$, the images under applications of $D$ and
$D^{-1}$ are written $a_n=D^n\Phi\in C_{NL}^\infty$ when $n\leq
0$, $a_n\in C_\downarrow^\infty$ when $n\geq1.$

Multiplication from the left by $a_n$ gives a linear operator
$C_{NL}^\infty\rightarrow C_{NL}^\infty,$ as do $D$ and $D^{-1}$.

The formulas \begin{eqnarray*} Da_1&+&a_1D=a_2 \\ Da_2&-&a_2D=a_3
\\  Da_3&+&a_3D=a_4, \\ & \vdots &
\end{eqnarray*} are identities in the ring of linear operators
$C_{NL}^\infty\rightarrow C_{NL}^\infty.$

The KdV element corresponding to $a_1=\Phi$ is $b_3=D^2b_1$ with
$b_1=a_5+3a_1a_2.$

Since the integration functional $C_I^\infty\rightarrow\Lambda$ is
identically zero on $DC_\downarrow^\infty\subset C_I^\infty,$ it
suffices to do computations in the quotient space
$C_I^\infty/DC_\downarrow^\infty.$ Thus for example
$b_1D^2\Phi=b_2D\Phi=-b_3\Phi$ for any $\Phi\in C_{NL}^\infty,$
because $D(b_1D\Phi),D^2(b_1\Phi)\in DC_\downarrow^\infty.$

Further, a function $h:C_\downarrow^\infty\rightarrow C_I^\infty$
gives a conserved quantity for Super KdV if $\delta
h=\frac{d}{dt}|^{t=0}h(a_1+tb_3)=0,$ as an element of the quotient
space.

\section{The first gradient}
The operators in the recursion are written as before as
\begin{eqnarray*} & & K=D^3+a_1I \\ & &
P=D^5+3a_1D^2+a_2D+2a_3I.\end{eqnarray*} With $0=\cdots
=\widetilde{\Gamma}_0=\widetilde{\Omega}_1$ and
$\widetilde{\Gamma}_2=\theta,$ the next step in the recursion is
$\widetilde{\Gamma}_4=D^{-2}K\widetilde{\Gamma}_2=D^{-2}a_1\theta.$

Then $b_3\widetilde{\Gamma}_4=-b_1a_1\theta.$

The function $h=a_1a_{-1}\theta$ sends $C_\downarrow^\infty$ into
itself, and its gradient is computed by
\begin{eqnarray*} \delta h &=& b_3a_{-1}\theta+a_1b_1\theta \\
&=& -b_1D^2a_{-1}\theta-b_1a_1\theta \\ &=& -2b_1a_1\theta.
\end{eqnarray*} The equality \[\frac{d}{dt}|^{t=0}h(a_1+tb_3)=2b_3\widetilde{\Gamma}_4(a_1)=0\]
in the quotient space $C_I^\infty/DC_\downarrow^\infty$ proves
that $\int h$ is a conserved quantity for the KdV equation.

\section{The second gradient} In general the recursion operator
taking gradient to gradient can be written as
\begin{eqnarray*}D^{-2}KD^{-2}P &=& D^4+D^{-2}L_2+D^{-2}L_3 \\ L_2 &=& -2a_1D^3+4a_2D^2-a_3D+2a_4I
\\ L_3 &=& 2a_1D^{-2}a_1D^2+a_1D^{-1}a_1D,   \end{eqnarray*} after the operator
identities $Da_1=a_2-a_1D$ and $D^2a_1=a_3+a_1D^2$ are taken into
account.

From $\widetilde{\Gamma}_4=D^{-2}a_1\theta,$ the recursion gives
the second gradient $\widetilde{\Gamma}_8$ as the sum of three
terms. An antigradient of $\widetilde{\Gamma}_8$ would satisfy
\begin{eqnarray*}\frac{d}{d\varepsilon}|^{\varepsilon=0}\widetilde{h}_8(a_1+\varepsilon
b_3)&=& b_3\widetilde{\Gamma}_8 \\ &=&
-b_1D^2\widetilde{\Gamma}_8.\end{eqnarray*} Since
$\widetilde{\Gamma}_8=(D^4+D^{-2}L_2+D^{-2}L_3  )
\widetilde{\Gamma}_4,$ we should examine
\[b_1D^2\widetilde{\Gamma}_8=b_1D^6\widetilde{\Gamma}_4
+b_1L_2\widetilde{\Gamma}_4+b_1L_3\widetilde{\Gamma}_4.
\] The first term is \[b_1D^6\widetilde{\Gamma}_4=b_1D^4a_1\theta=b_1a_5\theta. \]
An antigradient is given by \[h=\frac{1}{2}a_1a_3\theta\] because
\begin{eqnarray*}\delta h &=& \frac{1}{2}b_3a_3\theta+\frac{1}{2}a_1b_5\theta \\
&=& -\frac{1}{2}b_1D^2a_3\theta-\frac{1}{2}b_5a_1\theta \\ &=&
-\frac{1}{2}b_1a_5\theta-\frac{1}{2}b_1D^4a_1\theta \\ &=&
-b_1a_5\theta.
\end{eqnarray*}
Therefore the function $C_\downarrow^\infty\rightarrow
C_\downarrow^\infty$ given by $-h$ has the first term of
$\widetilde{\Gamma}_8$ as its gradient.

When the operator $L_2$ is applied to
$\widetilde{\Gamma}_4=D^{-2}a_1\theta,$ the result is
\[(2a_{-1}a_4-a_0a_3+2a_1a_2)\theta-a_{-1}a_3,
\] the second term in $D^2\widetilde{\Gamma}_8$.

Working with $h=a_{-1}a_1a_2\theta,$ we find that $\delta
h=x+y+z,$ with
\begin{eqnarray*}x &=& b_1a_1a_2\theta \\ y &=& a_{-1}b_3a_2\theta
\\ z &=& a_{-1}a_1b_4\theta.
\end{eqnarray*} Then \begin{eqnarray*}y &=& -b_3a_{-1}a_2\theta \\
&=& b_1D^2a_{-1}a_2\theta \\ &=&
b_1a_1a_2\theta+b_1a_{-1}a_4\theta,
\end{eqnarray*} while \begin{eqnarray*}z &=& -b_2D^2a_{-1}a_1\theta \\
&=& -b_2a_{-1}a_3\theta \\ &=& -b_1Da_{-1}a_3\theta \\ &=&
-b_1a_0a_3\theta+b_1a_{-1}a_4\theta-b_1a_{-1}a_3.
\end{eqnarray*} This gives \[\delta
h=b_1(2a_1a_2\theta-a_0a_3\theta+2a_{-1}a_4\theta)-b_1a_{-1}a_3=b_1L_2\widetilde{\Gamma}_4,\]
proving that $h=a_{-1}a_1a_2\theta$ has for its gradient the
second term in $\widetilde{\Gamma}_8.$

Finally, the operator $L_3$ is applied to
$\widetilde{\Gamma}_4=D^{-2}a_1\theta,$ giving just
\[a_1D^{-1}a_1D^{-1}a_1\theta.
\] Using
$D^{-1}a_1\theta=a_0\theta-a_{-1},Da_{-1}a_0=a_0^2-a_{-1}a_1,$ and
$Da_0^2\theta=2a_0a_1\theta+a_0^2,$ one can show that
\[2a_1D^{-1}a_1D^{-1}a_1\theta=a_0^2a_1\theta+2a_{-1}a_0a_1+a_1D^{-1}a_0^2,  \]
this being a constant multiple of the third term of
$D^2\widetilde{\Gamma}_8.$

An antigradient exists, and is a constant multiple of
\[h=a_0^4\theta-4a_{-1}a_0^3+3a_0^2D^{-1}a_0^2,  \] a function
$C_\downarrow^\infty\rightarrow C_{NL}^\infty.$ In order for it to
be integrable we need to show that $\frac{\partial}{\partial
\theta}h\in C_\downarrow^\infty,$ that is, $h\in C_I^\infty.$

It is easy to see that $\frac{\partial}{\partial \theta}a_0^N\in
C_\downarrow^\infty$ for all $N>0$. Remembering that
$C_\downarrow^\infty$ is an ideal in $C_{NL}^\infty$ we have
\[ \frac{\partial h}{\partial \theta}=a_0^4-4a_0^3\frac{\partial}{\partial \theta}a_{-1}
+3a_0^2\frac{\partial}{\partial \theta}D^{-1}a_0^2,   \] except
for a term in $C_\downarrow^\infty$.

Since $(\frac{\partial}{\partial
\theta}D^{-1}+D^{-1}\frac{\partial}{\partial \theta})\Phi=\Phi$
for all $\Phi\in C_{NL}^\infty,$ we obtain
\[\frac{\partial}{\partial \theta}a_{-1}=\frac{\partial}{\partial \theta}D^{-1}a_0=
a_0-D^{-1}\frac{\partial a_0}{\partial \theta}\equiv a_0
\mathrm{\:mod\:} C_\downarrow^\infty
\] while \[\frac{\partial}{\partial \theta}D^{-1}a_0^2=a_0^2-D^{-1}\frac{\partial}{\partial \theta}
a_0^2\equiv a_0^2\mathrm{\:mod\:}C_\downarrow^\infty,   \] because
$D^{-1}\frac{\partial}{\partial \theta}C_I^\infty\subset
C_\downarrow^\infty.$

This proves that $h$ takes its values in $C_I^\infty,$ as claimed,
because the powers of $a_0$ all cancel.

The gradients of the three terms of $h$ are now computed.

\begin{eqnarray*}\delta(a_0^4\theta) &=& 4a_0^3b_2\theta \\
&=& 4b_1Da_0^3\theta \\ &=& b_1(12a_0^2a_1\theta+4a_0^3).
\end{eqnarray*}

\begin{eqnarray*}\delta(a_{-1}a_0^3) &=& b_1a_0^3+3a_{-1}a_0^2b_2 \\
&=& b_1a_0^3+3b_1Da_{-1}a_0^2 \\ &=&
b_1a_0^3+3b_1(a_0^3-2a_{-1}a_0a_1)\\ &=&
b_1(4a_0^3-6a_{-1}a_0a_1).
\end{eqnarray*}

$\delta\frac{1}{2}(a_0^2D^{-1}a_0^2)=x+y$ in which
\begin{eqnarray*}x &=& a_0b_2D^{-1}a_0^2 \\
&=& b_1Da_0D^{-1}a_0^2 \\ &=& b_1(a_1D^{-1}a_0^2+a_0^3),
\end{eqnarray*} while $y=a_0^2D^{-1}a_0b_2.$

Using $D^{-1}a_0b_2=a_0b_1-D^{-1}a_1b_1$ and
$a_1b_1=a_1(a_5+3a_1a_2)=D^2a_1a_3,$ we obtain
\[y=b_1a_0^3-a_0^2Da_1a_3.\] But since $(D^{-1}a_0^2)(Da_1a_3)\in
C_\downarrow^\infty$, we can apply $D$, obtaining
\[a_0^2Da_1a_3=(D^{-1}a_0^2)D^2a_1a_3=(D^{-1}a_0^2)a_1b_1,      \]
mod $DC_\downarrow^\infty.$

This gives \begin{eqnarray*}y &=& b_1a_0^3-(D^{-1}a_0^2)a_1b_1 \\
&=& b_1(a_0^3+a_1D^{-1}a_0^2) ,
\end{eqnarray*} and therefore
\[\delta(a_0^2D^{-1}a_0^2)=b_1(4a_0^3+4a_1D^{-1}a_0^2).   \]
Taken in combination with
\begin{eqnarray*}\delta(a_0^4\theta) &=& b_1(4a_0^3+12a_0^2a_1\theta) \\
-4\delta(a_{-1}a_0^3) &=& b_1(-16a_0^3+24a_{-1}a_0a_1) ,\\
3\delta(a_0^2D^{-1}a_0^2)&=&b_1(12a_0^3+12a_1D^{-1}a_0^2)
\end{eqnarray*} this completes the proof that

\begin{eqnarray*}\delta h &=& b_1(12a_0^2a_1\theta+24a_{-1}a_0a_1+12a_1D^{-1}a_0^2) \\
&=&
b_1(24a_1D^{-1}a_1D^{-1}a_1\theta)=24b_1L_3\widetilde{\Gamma}_4 .
\end{eqnarray*} To sum up: the superfield
$\widetilde{\Gamma}_8(a_1)$ produced from the initial value
$\widetilde{\Gamma}_2=\theta$ and the subsequent constructions
given by the recursive algorithm has been shown to appear in an
equation \[\frac{d}{dt}|^{t=0}\int\widetilde{h}_8(a_1+tb_3)=\int
b_3\widetilde{\Gamma}_8(a_1)
\] for a certain $\widetilde{h}_8:C_\downarrow^\infty\rightarrow
C_I^\infty.$

This shows that $a_1\rightarrow\int\widetilde{h}_8(a_1),$ a
function $C_\downarrow^\infty\rightarrow\Lambda,$ is a nonlocal
conserved quantity for the SUSY KdV equation.

\section{Conclusions} We presented a complete proof of the
gradient recursion algorithm for the $N=1$ SKdV system. We
introduced the precise ring of superfields where the non-local
gradients and conserved quantities appear. All the local and
non-local hierarchy of the $N=1$ SKdV is obtained from the
gradient recursion algorithm. In particular we found new non-local
conserved quantities of the $N=1$ SKdV equation. These new
conserved quantities are bosonic in contrast to the already known
fermionic non-local conserved quantities. They were constructed
step by step using the recursive gradient algorithm. That suggests
that there might exist a new non-local conserved quantity of the
Super Gardner equation (S. Andrea, A. Restuccia and A. Sotomayor,
work in progress).

The recursive gradient approach may also be extended for $N=2$
SKdV equations \cite{MMM}, we expect to report on this shortly.

\newpage \textbf{Acknowledgments} The work of A.R.
was supported by PROSUL under contract CNPq 490134/2006-8 and
Decanato de Investigaci\'on y Desarrollo(DID USB), Proyecto G11.


\begin{thebibliography}{}
\bibitem{Witten}{E. Witten, Surveys in Differential Geometry 1,
243-310 (1991).}
\bibitem{Restuccia1}{I. Martin, A. Restuccia and R. S. Torrealba, Nucl. Phys. B521, 117-128 (1998). }
\bibitem{Restuccia2}{L. Boulton, M. Garcia del Moral and A.
Restuccia, Nucl. Phys. B671, 343-358 (2003).}
\bibitem{Restuccia3}{J. Bellorin and A. Restuccia, Nucl. Phys.
B737, 190-208 (2006).}
\bibitem{Restuccia4}{I. Martin and A. Restuccia, Nucl. Phys. B622, 240-256 (2002). }
\bibitem{Manin}{Yu. I. Manin and A. O. Radul, Commun. Math. Phys.
98, 65 (1985).}
\bibitem{MM}{P. Mathieu, Lett. Math. Phys. 16, 199 (1988).}
\bibitem{M}{ P. Mathieu, J. Math. Phys. 29, 2499
(1988).}
\bibitem{Mathieus}{P. Mathieu, ``Open problems for the super KdV equations",
math-ph/0005007.}
\bibitem{Andrea2}{S. Andrea, A. Restuccia, A. Sotomayor, J. Math.
Phys. 45, 1715 (2004).}
\bibitem{Andrea3}{S. Andrea, A. Restuccia, A. Sotomayor, J. Math.
Phys. 46, 103517 (2005).}
\bibitem{Gardner et al}{ R. M. Miura, C. S. Gardner, and M. D. Kruskal,
J. Math. Phys. 9, 1204 (1968).}
\bibitem{Kiselev1}{A. V. Kiselev and T. Wolf, SIGMA 2, 030, (2006).}
\bibitem{Kiselev2}{A. Karasu and A. Kiselev, Math. Gen. 39, 11453 (2006).}
\bibitem{Andrea1}{S. Andrea, A. Restuccia, A. Sotomayor, J. Math.
Phys. 42, 2625 (2001).}
\bibitem{Kersten1}{P. H. M. Kersten, Phys. Lett. A 134, 25 (1988).}
\bibitem{Dargis}{P. Dargis and P. Mathieu, Phys. Lett. A 176,
67-74 (1993).}
\bibitem{MMM}{C. A. Laberge and P. Mathieu, Phys. Lett. B215, 718
(1988).}





\end{thebibliography}
\end{document}